\documentstyle[12pt]{article}
\newfont{\ffont}{msym10}                        
\newcommand{\beq}{\begin{equation}}             
\newcommand{\eeq}{\end{equation}}               
\newcommand{\bqry}{\begin{eqnarray}}            
\newcommand{\eqry}{\end{eqnarray}}              
\newcommand{\bqryn}{\begin{eqnarray*}}          
\newcommand{\eqryn}{\end{eqnarray*}}            
\newcommand{\NL}{\nonumber \\}                  
\newcommand{\preprint}[1]{\begin{table}[t]      
            \begin{flushright}                  
            \begin{large}{#1}\end{large}        
            \end{flushright}                    
            \end{table}}                        
\newcommand{\PD}[2]                             
    {\frac{\partial^{#2}}{\partial #1^{#2}}}    
\catcode`\@=11 \@addtoreset{equation}{section}  
\renewcommand{\theequation}                     
         {\arabic{section}.\arabic{equation}}   
\begin{document}
\preprint{TAUP-2554-95 \\  }
\title{Generalized Boltzmann Equation \\ in a Manifestly Covariant \\
Relativistic Statistical Mechanics}
\author{\\ L. Burakovsky\thanks {Bitnet: BURAKOV@TAUNIVM.TAU.AC.IL} \
and L.P. Horwitz\thanks {Bitnet: HORWITZ@TAUNIVM.TAU.AC.IL. Also at
Department of Physics, Bar-Ilan University, Ramat-Gan, Israel  } \\ \ }
\date{School of Physics and Astronomy \\ Raymond and Beverly Sackler
Faculty of Exact Sciences \\ Tel-Aviv University, Tel-Aviv 69978, Israel}
\maketitle
\begin{abstract}
We consider the relativistic statistical mechanics of an ensemble of
$N$ events with motion in space-time parametrized by an invariant
``historical time'' $\tau .$ We generalize the approach of Yang and Yao,
based on the Wigner distribution functions and the Bogoliubov hypotheses,
to find the approximate dynamical equation for the kinetic state of any
nonequilibrium system to the relativistic case, and obtain a manifestly
covariant Boltzmann-type equation which is a relativistic generalization
of the Boltzmann-Uehling-Uhlenbeck (BUU) equation for indistinguishable
particles. This equation is then used to prove the $H$-theorem for
evolution in $\tau .$ In the equilibrium limit, the covariant forms of
the standard statistical mechanical distributions are obtained. We
introduce two-body interactions by means of the direct action potential
$V(q),$ where $q$ is an invariant distance in the Minkowski space-time.
The two-body correlations are taken to have the support in a relative $O(
2,1)$-invariant subregion of the full spacelike region. The expressions
for the energy density and pressure are obtained and shown to have the
same forms (in terms of an invariant distance parameter) as those of the
nonrelativistic theory and to provide the correct nonrelativistic limit.
\end{abstract}
\bigskip
{\it Key words:} special relativity, relativistic Boltzmann equation,
relativistic Maxwell-Boltzmann/Bose-Einstein/Fermi-Dirac, mass
distribution

PACS: 03.30.+p, 05.20.Gg, 05.30.Ch, 98.20.--d
\bigskip

\section{Introduction}
This paper continues a series of works on relativistic kinetic theory of
an $N$-body system \cite{HSP}--\cite{hadr}
within the framework of a manifestly covariant
mechanics \cite{HP}, both for the classical theory and the corresponding
relativistic quantum theory. In this framework, for the classical
case, the covariant dynamical evolution of a system of $N$ particles is
governed by equations of motion that are of the form of Hamilton
equations for the motion of $N\;events$ which generate the particle
space-time trajectories (world lines). These events are considered as the
fundamental dynamical objects of the theory and characterized by their
positions $q^{\mu }=(ct,{\bf q})$ and energy-momenta $p^{\mu }=(E/c,{\bf
p})$ in an $8N$-dimensional phase space. The motion is parametrized by a
continuous Poincar\'{e}-invariant parameter $\tau $ \cite{HP} called the
``historical time''. For the quantum case, the covariant dynamical
evolution of $N$ particles is governed by a generalized Schr\"{o}dinger
equation for the wave function $\psi _{\tau }(q_1,q_2,...,q_N)\in
L^2(R^{4N}),$ with measure $dq_1dq_2\cdots dq_N\equiv d^{4N}q,$
describing the distribution of events ${q_i\equiv q_i^{\mu },\;\mu =0,1,2
,3;\;i=1,2,\ldots ,N}.$ The collection of events (called
``concatenation'' \cite{AHL}) along each world line  corresponds to a
$particle$ in the usual sense; e.g., the Maxwell conserved current is an
integral over the history of the charged event \cite{Jack}. Hence the
evolution of the state of the $N$-event system describes $a\;posteriori$
the history in space and time of an $N$-particle system.

The evolution of the system is assumed to be governed by Hamiltonian-type
equations with a Lorentz-invariant scalar function, the relativistic
dynamical function of the variables $(q_i,p_i)$ specifying the state of
each particle $i.$ In the simplest case of a free particle, for which the
world line is generated by a free event, the relativistic dynamical
function (generalized Hamiltonian) is $$K_0=\frac{p^\mu p_\mu }{2M},$$
where we use the metric $g^{\mu \nu }=(-,+,+,+),$ and $M$ is a given
fixed parameter (an intrinsic property of the event), with the dimension
of mass.

The Hamilton equations $$\frac{dq^\mu }{d\tau }=\frac{\partial K}
{\partial p_\mu },\;\;\;\frac{dp^\mu }{d\tau }=-\frac{\partial K}
{\partial q_\mu }$$ yield, in this case, $$\frac{dq^\mu }{d\tau }=\frac{p
^\mu }{M},\;\;\;\frac{dp^\mu }{d\tau }=0.$$ Eliminating $d\tau ,$ one
finds $$\frac{d{\bf q}}{dt}=\frac{{\bf p}}{E}c^2,$$ as required for the
motion of a free relativistic particle. It then follows that, for a free
motion, the proper time interval squared, divided by $d\tau ^2,$ is
$$\frac{dq^\mu }{d\tau }\frac{dq_\mu }{d\tau }=\frac{p^\mu p_\mu }
{M^2}.$$ For $$K_0=-\frac{M}{2},$$ corresponding to the ``mass-shell''
value $$p^\mu p_\mu =-M^2c^2,$$ it follows that $$c^2dt^2-d{\bf q}^2=c^2d
\tau ^2.$$ In the more general case in which $$K=K_0+V,$$ where $V$ is,
for example, a function of $q,\;\;p^2\equiv p^\mu p_\mu $ may vary from
point to point along the trajectory. Hence, in general, the proper time
interval does $not$ correspond to $d\tau .$

For a system of $N$ interacting events (and hence, particles) one takes
\cite{HP}
\beq
K=\sum _i\frac{p^\mu _ip_{i\mu }}{2M}+V(q_1,q_2,\ldots ,q_N),
\eeq
where all of the events are put, for simplicity, to have equal mass
parameters, and we write $q_i,$ for brevity, for the four-vector.
The Hamilton equations are
$$\frac{dq^\mu _i}{d\tau }=\frac{\partial K}{\partial p_{i\mu }}=\frac{p_
i^\mu }{M},$$
\beq
\frac{dp^\mu _i}{d\tau }=-\frac{\partial K}{\partial q_{i\mu }}=
-\frac{\partial V}{\partial q_{i\mu }}.
\eeq
These equations are precisely of the same form as those of
nonrelativistic Hamilton point mechanics, but in a space of $8N$
dimensions instead of $6N.$ The fundamental theorems of mechanics, such
as the Liouville theorem \cite{HSS}, the theory of
canonical transformations and Hamilton-Jacobi theory, follow in the same
way, with the manifold of space-time replacing that of space, and
energy-momentum replacing the momentum. It is fundamental to this
structure that there is a single universal evolution parameter $\tau $
which plays the role of the Galilean time.

In the quantum theory, the generalized Schr\"{o}dinger equation
\beq
i\hbar \frac{\partial }{\partial \tau }\psi _\tau (q_1,q_2,\ldots ,q_N)=K
\psi _\tau (q_1,q_2,\ldots ,q_N),
\eeq
with, for example, a $K$ of the form (1.1), describes the evolution of
the $N$-body wave function $\psi _\tau (q_1,q_2,\ldots ,q_N).$ To
illustrate the meaning of this wave function, consider the case of a
single free event. In this case, (1.3) has the formal solution $$\psi _
\tau (q)=(e^{-iK_0\tau /\hbar }\psi _0)(q)$$ for the evolution of the
free wave packet. Let us represent $\psi _\tau (q)$ by its Fourier
transform, in the energy-momentum space: $$\psi _\tau (q)=\frac{1}{(2\pi
\hbar )^2}\int d^4p\;e^{-ip^2\tau /2M\hbar }e^{ip\cdot q/\hbar }\psi _
0(p),$$ where $p^2\equiv p^\mu p_\mu,\;p\cdot q\equiv p^\mu q_\mu ,$ and
$\psi _0(p)$ corresponds to the initial state. Applying the Ehrenfest
arguments of stationary phase to obtain the principal contribution to
$\psi _\tau (q)$ for a wave packet centered at $p_c^\mu ,$ we find $$q_c^
\mu =\frac{p_c^\mu }{M}\tau ,$$ consistent with the classical equations
(1.2). Therefore, the central peak of the wave packet moves along the
classical trajectory of an event, i.e., the classical world line.

The wave functions have a local interpretation, i.e., $\vert \psi _\tau
(q)\vert ^2d^4q$ is the probability to find an event at the space-time
point $q^\mu $ in space-time volume $d^4q.$ Localization in space, as
well as in time, can be shown by applying arguments given in ref. \cite
{AH}.

Horwitz, Schieve and Piron \cite{HSP} have constructed equilibrium
classical and quantum Gibbs ensembles. They found that the grand
partition function in the rest frame of the system is given by
\beq
\ln Z(\beta ,V,\mu ,\mu _K)=e^{\beta \mu }\int \frac{d^4pd^4q}{(2\pi )^4}
e^{-\beta E}e^{\beta \mu _K\frac{m^2}{2M}},\;\;\beta =\frac{1}{k_BT}.
\eeq
In addition to the usual chemical potential $\mu $ in the grand canonical
ensemble, there is a new potential $\mu _K$ corresponding to the mass
degree of freedom of relativistic systems.

Horwitz, Shashoua and Schieve \cite{HSS} have shown that in the framework
of the manifestly covariant mechanics which we discuss here, covariant
Weyl transforms exist for observables, and therefore covariant
relativistic Wigner functions \cite{DK} can be constructed. In this way
they derived a manifestly covariant relativistic generalization of the
BBGKY hierarchy for the $s$-particle relativistic Wigner functions. By
approximating the effect of correlation of second and higher order by
two-body collision terms (using the cross-sections defined in ref.
\cite{HL}), as in the usual nonrelativistic Boltzmann theory, they
obtained a manifestly covariant Boltzmann equation (for non-identical
events). This equation was used to prove the $H$-theorem for evolution
in $\tau .$ In the equilibrium limit, a covariant form of the
Maxwell-Boltzmann distribution,
\beq
f^{(0)}(q,p)=e^{A(q)(p-p_c)^2},
\eeq
was obtained. Since this distribution
is the distribution of the $4$-momenta of the events, $m^2=-p^2=-p^{\mu }
p_{\mu }$ is a random variable in a relativistic ensemble. In order to
obtain a simple analytic result the authors restricted themselves to a
narrow mass shell $p^{2}=-m^{2}\cong -M^{2}.$ The results obtained in
this approximation are in agreement with the well-known results of Synge
\cite{Syn} for an on-shell relativistic kinetic theory.

In ref. \cite{di} the equilibrium Maxwell-Boltzmann distribution (1.5)
was considered for the whole range of $m,$ to obtain the corresponding
equilibrium relativistic $mass$ distribution. Its low-temperature and
nonrelativistic limits were investigated and shown to yield results in
agreement with nonrelativistic statistical mechanics \cite{galim}.

In the present paper we study the case of indistinguishable events.
In contrast to the approach of Horwitz, Shashoua and Schieve \cite{HSS},
we choose another approach initiated by Yang and Yao \cite{YY} in the
nonrelativistic case, which is based on the Wigner distribution
functions and the Bogoliubov hypotheses to find approximate dynamical
equation for the kinetic state of any nonequilibrium system \cite{Bog}.
Kinetic equation that we obtain, which represents a relativistic
generalization of the Boltzmann-Uehling-Uhlenbeck (BUU) equation
\cite{BUU} for indistinguishable particles, and can be easily generalized
to include the non-identical case as well. The generalized Boltzmann
equation obtained in this way is then used to prove the $H$-theorem for
evolution in $\tau .$ In the equilibrium limit, the covariant forms of
the Bose-Einstein/Fermi-Dirac/Maxwell-Boltzmann distributions are
obtained, which, as considered for the whole range of $m,$ provide the
corresponding equilibrium relativistic mass distributions. The
relativistic mass distributions are studied in the identical particle
case in \cite{ind}, and their possible consequences for high energy
physics and cosmology are considered, respectively, in \cite{hadr} and
\cite{therm}.

We introduce two-body interactions by taking the support of mutual
correlations for any two events to be in a relative $O(2,1)$-invariant
subregion of the full spacelike region, as done in the solution of the
two-body bound state problem \cite{AH1,AH2}, and for the extraction of
the partial wave expansion from the relativistic scattering amplitude
\cite{AH3}. We then calculate the expressions for the energy density and
pressure of an interacting gas, and show that they have the same form (in
terms of an invariant distance parameter) as those of the nonrelativistic
theory and provide the correct nonrelativistic limit.

\section{Relativistic $N$-body system}
The evolution in $\tau $ of an $N$-body system is determined by the
Liouville-von Neumann equation for the $N$-body density matrix $\rho $
(we use the system of units in which $\hbar =c=k_B=1,$ unless other units
are specified):
\beq
i\frac{\partial \rho }{\partial \tau }=[K,\rho ],
\eeq
where $K$ is the total $N$-body Hamiltonian, here taken to be
\beq
K=\sum _{i=1}^NK_i^{(0)}+\sum _{1=i<j}^NV_{i,j},
\eeq
where $$K_i^{(0)}=\frac{p_i^\mu p_{i\mu }}{2M}$$ and $$V_{i,j}=V\left( q_
i-q_j\right) ,\;\;\;q_i-q_j\equiv \sqrt {(q_i^\mu -q_j^\mu )(q_{i\mu }-q_
{j\mu })}$$ is a two-body interaction potential. In order to obtain the
BBGKY hierarchy, one introduces the $(n)$-body density matrices, as
follows:
\beq
\rho ^{(n)}_{1,2,\ldots ,n}=\frac{N!}{(N-n)!}Tr_{(n+1,\ldots ,N)}\rho ,
\eeq
\beq
Tr_{(1,2,\ldots ,n)}\rho ^{(n)}_{1,2,\ldots ,n}=\frac{N!}{(N-n)!},
\eeq
and, by taking the appropriate traces in Eq. (2.1), obtains \cite{BM}
\bqry
i\frac{\partial \rho ^{(n)}_{1,2,\ldots ,n}}{\partial \tau } & = &
\sum _{i=1}^n\;[K_i^{(0)},\rho ^{(n)}_{1,2,\ldots ,n}]\;+\sum _{1=i<j}^
n[V_{i,j},\rho ^{(n)}_{1,2,\ldots ,n}] \NL
 &   & +\;Tr_{(n+1)}\sum _{i=1}^n\;[V_{i,n+1},\rho ^{(n+1)}_{
1,2,\ldots ,n+1}].
\eqry
This set of equations is equivalent to (2.1).

In what follows, we shall use the simplified notation: $\rho _i\equiv
\rho _i^{(1)},\;\rho _{i,j}\equiv \rho _{i,j}^{(2)},$ etc., so that the
latter equation can be rewritten as
\beq
i\frac{\partial \rho _n}{\partial \tau }=\sum _{i=1}^n\;[K_i^{(0)},
\rho_n]+\!\sum _{1=i<j}^n[V_{i,j},\rho _n]+Tr_{(n+1)}\sum _{i=1}^n\;
[V_{i,n+1},\rho _{n+1}].
\eeq
It is convenient to introduce directly the symmetry requirements on the
function $\rho _n$ by means of
\beq
\rho _n=S_nF_n,
\eeq
where $S_n$ is a symmetrization/antisymmetrization operator defined by
\beq
S_n=\prod _{i=2}^n\left( 1\pm \sum _{j=1}^{i-1}P_{i,j}\right) .
\eeq
Here $P_{i,j}$ denotes the permutation operator. Since $S_n$ satisfies
the relation
\beq
S_{n+1}=S_n\left( 1\pm \sum _{i=1}^nP_{i,n+1}\right)
\eeq
and commutes with the operators $K_i$ and $V_{i,j},$ one can substitute
(2.7) into (2.6) and obtain the equation
\bqry
i\frac{\partial F_n}{\partial \tau } & = & \sum _{i=1}^n\;[K_i^{(0)},
F_n]\;+\sum _{1=i<j}^n[V_{i,j},F_n]\;+\;
Tr_{(n+1)}\sum _{i=1}^n\;[V_{i,n+1},F_{n+1}] \NL
  &   & \pm \;\;
Tr_{(n+1)}\sum _{i=1}^n\;[V_{i,n+1},\sum _{i=1}^nP_{i,n+1}F_{n+1}].
\eqry
Now we introduce the Wigner distribution functions \cite{DK},
\beq
f_s(q_s,p_s,\tau )=\frac{1}{(2\pi )^{4s}}\int dr_s\;F_s(q^{'}_s,q^{''}_s,
\tau )e^{ip_s\cdot r_s},
\eeq
\beq
F_s(q^{'}_s,q^{''}_s,\tau )=\int dp_s\;f_s(q_s,p_s,\tau )e^{
-ip_s\cdot r_s},
\eeq
where $$q^{'}_s=q_s-\frac12r_s,\;\;\;q^{''}_s=q_s+\frac12r_s,$$ and
$q_s\equiv (q_1,q_2,\ldots ,q_s),$ $p_s\equiv (p_1,p_2,\ldots ,p_s),$
$p_s\cdot r_s\equiv \sum _{i=1}^sp^\mu _ip_{i\mu },$ $dr_s\equiv dr_1dr_
2\cdots dr_s.$ One may substitute (2.12) into (2.10) and obtain the
quantum BBGKY hierarchy of the Wigner distribution functions $f_s=f_s(x_
s,\tau ),\;\;x_s=(q_s,p_s),$ as
\bqry
\frac{\partial f_s}{\partial \tau } & + & \sum _{j=1}^s\frac{p_j}{M}
\frac{\partial f_s}{\partial q_j}\;\;+\;\;i\sum _{j<k}^s\left( e^{i
\theta _{j,k}/2}-e^{-i\theta _{j,k}/2}\right) f_s \NL
 & + & i\sum _{j=1}^s\int dx_{s+1}\;\left( e^{i\theta _{j,s+1}/2}-e^{-i
\theta _{j,s+1}/2}\right) f_{s+1} \NL
 & \pm & i\sum _{j=1}^s\int dx_{s+1}\;\left( e^{i\theta _{j,s+1}/2}-e^{-i
\theta _{j,s+1}/2}\right) P_{j,s+1}f_{s+1}\;=\;0.
\eqry
Here $dx_s\equiv dq_sdp_s=d^4q_1\cdots d^4q_sd^4p_1\cdots d^4p_s,$ and
the operators
$\theta _{ij}$ and $\theta _{j,s+1}$ are represented as follows,
\beq
\theta _{ij}=\frac{\partial V_{ij}}{\partial q_i}\left( \frac{\partial }
{\partial p_i}-\frac{\partial }{\partial p_j}\right) ,\;\;\;\theta _{j,s+
1}=\frac{\partial V_{ij}}{\partial q_i}\frac{\partial }{\partial p_j}.
\eeq
For $s=1$ and 2, one finds
\bqry
\frac{\partial f_1}{\partial \tau } & + & \frac{p_1}{M}\frac{\partial
f_1}{\partial q_1}\;\;+\;\;i\int dx_2\;\left( e^{i\theta _{1,2}/2}-e^{-i
\theta _{1,2}/2}\right) f_2 \NL
 & \pm & i\int dx_2\;\left( e^{i\theta _{1,2}/2}-e^{-i
\theta _{1,2}/2}\right) P_{1,2}f_2\;=\;0, \\
\frac{\partial f_2}{\partial \tau } & + & \sum _{j=1}^2\frac{p_j}{M}
\frac{\partial f_2}{\partial q_j}\;+\;i\left( e^{i\theta _{
1,2}/2}-e^{-i\theta _{1,2}/2}\right) f_2 \NL
 & + & i\sum _{j=1}^2\int dx_3\;\left( e^{i\theta _{j,3}/2}-e^{-i
\theta _{j,3}/2}\right) f_3 \NL
 & \pm & i\sum _{j=1}^2\int dx_3\;\left( e^{i\theta _{j,3}/2}-e^{-i
\theta _{j,3}/2}\right) P_{j,3}f_3\;=\;0.
\eqry
Equations (2.15) and (2.16) are exact. Since $f_2$ depends on $f_3,$
accurate solution of the hierarchy is impossible. One has, therefore, to
apply some approximated approach. One of such approaches is the
Bogoliubov one \cite{Bog}, which we shall apply in the present
consideration.

According to the Bogoliubov hypotheses \cite{Bog},

1) It is possibe to find a kinetic state of any non-equilibrium system,
provided that the average interval between two subsequent collisions is
much longer than the duration of the collision. In this kinetic state,
\beq
f_s(x_1,\ldots ,x_s;\tau )=f_s(x_1,\ldots ,x_s|f_1),
\eeq
\beq
\frac{\partial f_1}{\partial \tau }=A(x|f_1).
\eeq

2) There are no correlations in the initial state of a system. One can
introduce the displacement operator,
\beq
{\cal P}^s_\tau f_s(x^0_1,\ldots ,x_s^0)=f_s(x_1,\ldots ,x_s),
\eeq
where $x_1^0,\ldots ,x_s^0$ are the values of each $x$ at $\tau =0,$ and
$x_1,\ldots ,x_s$ are their values at $\tau .$ The non-correlative
condition at the initial state implies
\beq
{\cal P}^s_{-\tau }\left[ f_s(x_1,\ldots ,x_s)-\prod _{1\leq j\leq s}
f_1(x_j)\right] \rightarrow 0.
\eeq
Starting from the Bogoliubov hypotheses, it is possible to derive a
kinetic equation.

Although the invariant interaction potential has infinite support in
space-time, since it depends on $({\bf q}_1-{\bf q}_2)^2-c^2(t_1-t_2)^2,$
its long-range part is necessary close to the light cone. It has been
shown \cite{HS}, that wave operators exist in scattering theory if the
support of the wave function does not extend to zero mass. The space-time
volume $v$ of the effective interaction is therefore bounded. We shall
assume here that it may be taken to be small, as in the first hypothesis
of Bogoliubov. One can, therefore, write
\beq
\frac{\partial f_1}{\partial \tau }=A^0(x|f_1)+vA^1(x|f_1)=\ldots ,
\eeq
\beq
f_s=f_s^0+vf_s^1+v^2f_s^2+\ldots .
\eeq
In the first-order approximation, one sets
\beq
f_2\cong f_2^0\cong f_1(1)f_1(2)
\eeq
(henceforth we use the notation $1\equiv (x_1;\tau ),\;2\equiv (x_2;\tau
),$ etc.) and finds from (2.15)
\bqry
\frac{\partial f_1(1)}{\partial \tau } & + & \frac{p_1}{M}\frac{
\partial f_1(1)}{\partial q_1}\;\;+\;\;i\int dx_2\;\left( e^{i\theta
_{1,2}/2}-e^{-i\theta _{1,2}/2}\right) f_1(1)f_1(2) \NL
 & \pm & i\int dx_2\;\left( e^{i\theta _{1,2}/2}-e^{-i
\theta _{1,2}/2}\right) P_{1,2}f_1(1)f_1(2)\;\;=\;\;0.
\eqry
This self-consistent equation is a relativistic generalization of the
quantum Vlasov equation \cite{YY}.

In the second-order approximation, one writes a formal solution,
\beq
f_s(x_1,\ldots ,x_s|f_1)=\sum _{i<j\leq s}g(x_i,x_j)\prod _{
k\neq i\neq j}f_1(k),
\eeq
where
\beq
g(x_i,x_j)=f_2^1(x_i,x_j|f_1)
\eeq
is a two-body correlation function, whose boundary condition is
\beq
\lim _{\tau \rightarrow \infty }{\cal P}_{-\tau }^{(2)}g(x_i,x_j)=0.
\eeq
Eq. (2.25) means that $s$-body effects are correlated by two-body
effects. One can write
\bqry
\frac{\partial f_2}{\partial \tau }\;=\;\frac{\partial f_2}{\partial
f_1}\frac{\partial f_1}{\partial \tau } & \approx  & \left( \frac{
\partial f_2^0}{\partial f_1}+v\frac{\partial f_2^1}{\partial f_1}\right)
\left[ A^0(x|f_1)+vA^1(x|f_1)\right]   \NL
 & \approx  & D_0f_2^0\;+\;v\left[ D_0\;g(x_1,x_2)+D_1f_2^0\right] ,
\eqry
where $$D_0\equiv A^0\frac{\partial }{\partial f_1},\;\;\;
D_1\equiv A^1\frac{\partial }{\partial f_1}.$$ One now uses Eqs.
(2.16),(2.25) and obtains
\bqry
D_0\;g(x_1,x_2) & + & \sum _{j=1}^2\frac{p_j}{M}\frac{\partial }{\partial
q_j}g(x_1,x_2)\;\;+\;\;i\sum_{j=1}^2\left( e^{i\eta _j/2}-e^{-i\eta _j/2}
\right) g(x_1,x_2) \NL
= & - & i\left( e^{i\theta ^{'}_{1,2}/2}-e^{-i\theta ^{'}_{1,2}/2}\right)
f_1(1)f_1(2)\;\;-\;\;i\int dx_3\left( e^{i\theta ^{'}_{1,3}/2}-e^{-i
\theta ^{'}_{1,3}/2}\right)  \NL
 &  & \times \;g(x_2,x_3)f_1(1)\;\;-\;\;i\int dx_3\left( e^{i\theta
^{'}_{2,3}/2}-e^{-i\theta ^{'}_{2,3}/2}\right) f_1(2)g(x_1,x_3) \NL
 & \mp & i\int dx_3\left( e^{i\theta ^{'}_{1,3}/2}-e^{-i\theta ^{'}_{1,
 3}/2}\right) f_1(1)f_1(2)f_1(3) \NL
 & \mp & i\int dx_3\left( e^{i\theta ^{'}_{2,3}/2}-e^{-i\theta ^{'}_{2,
 3}/2}\right) f_1(1)f_1(2)f_1(3).
\eqry
Once $g(x_1,x_2)$ is known, one can obtain
the two-order-approximated equation for $f_1:$
\bqry
\frac{\partial f_1(1)}{\partial \tau } & + & \frac{p_1}{M}\frac{
\partial f_1(1)}{\partial q_1}\;\;+\;\;i\left( e^{i\eta _1/2}-e^{-i\eta
_1/2}\right) f_1(1)\;\;+\;\;i\int dx_2\;\left( e^{i\theta
_{1,2}^{'}/2}-e^{-i\theta _{1,2}^{'}/2}\right)  \NL
 &  & \times \;g(x_1,x_2)\;\;\pm \;\;i\int dx_2\;\left( e^{i\theta
_{1,2}^{'}/2}-e^{-i\theta _{1,2}^{'}/2}\right) f_1(1)f_1(2)\;\;=\;\;0.
\eqry
Here $$\theta ^{'}_{1,2}=\frac{1}{v}\theta _{1,2},\;\;\theta ^{'}_{1,3}=
\frac{1}{v}\theta _{1,3},\;\;\eta _1=\frac{\partial U_1}{\partial q_1}
\frac{\partial }{\partial p_1},$$ and
\beq
U_1(q_1,\tau )=\frac{1}{v}\int dx_2\;f_1(2)V(q_1-q_2)
\eeq
is the mean-field potential.
In general, it is very difficult to obtain simultaneously solutions of
Eqs. (2.29) and (2.30). In the following section we show how Eq. (2.29)
can be solved for a quasihomogeneous system.

\subsection{Quasihomogeneous system}
The condition on a quasihomogeneous system is
\beq
g(x_1,x_2)=g(q_1-q_2,p_1,p_2)\equiv g(q,p_1,p_2),
\eeq
i.e., the correlation function depends only on the relative coordinates.
In this case, one obtains a formal solution for $g(q,p_1,p_2)$ by means
of the displacement techniques, as follows:
\bqry
g(q,p_1,p_2) & = & \int _0^\infty d\tau \;\left[ i\left\{ \left( e^{
\frac{i}{2}\frac{\partial }{\partial q}(\frac{\partial }{\partial p_1}-
\frac{\partial }{\partial p_2})}-e^{-\frac{i}{2}\frac{\partial }{\partial
q}(\frac{\partial }{\partial p_1}-\frac{\partial }{\partial p_2})}\right)
V\left( q-\frac{p_1-p_2}{M}\tau \right) \right\} \right.  \NL
 &  & \times f_1(1)f_1(2)\;+\;i\int dq^{'}dp_3\;\left(
e^{\frac{i}{2}\frac{\partial }{\partial q}\frac{\partial }{\partial
p_1}}-e^{-\frac{i}{2}\frac{\partial }{\partial q}\frac{\partial }{
\partial p_1}}\right) V\left( q-q^{'} \right.  \NL
 &  & \left. -\;\frac{p_1-p_2}{M}\tau \right) \times \left(
g(q^{'},p_2,p_3)f_1(1)\;\pm \;f_1(1)f_1(2)f_1(3)\right)  \NL
 &  & \pm \;i\int dq^{'}dp_3\;\left(
e^{\frac{i}{2}\frac{\partial }{\partial q}\frac{\partial }{\partial
p_2}}-e^{-\frac{i}{2}\frac{\partial }{\partial q}\frac{\partial }{
\partial p_2}}\right) V\left( q-q^{'}-\frac{p_1-p_2}{M}\tau \right)  \NL
 &  & \left. \times \left( g(q^{'},p_1,p_3)f_1(2)
\;\pm \;f_1(1)f_1(2)f_1(3)\right) \right]  \NL
 & = & i\int d\tau \;\left[ \left( e^{i\theta ^{'}_{1,2}/2}-e^{-i\theta
^{'}_{1,2}/2}\right) f_1(1)f_1(2)\right.  \NL
 &  & +\;\int dx_3\;\left( e^{i\theta ^{'}_{1,3}/2}-e^{-i\theta ^{'}_{1,
3}/2}\right) \times \Big( g(x_2,x_3)f_1(1)\;\pm
\;f_1(1)f_1(2)f_1(3)\Big)  \NL
 &  & \left. \pm \;\int dx_3\;\left( e^{i\theta ^{'}_{2,3}/2}-e^{-i\theta
^{'}_{2,3}/2}\right) \times \Big( g(x_1,x_3)f_1(2)\;\pm
\;f_1(1)f_1(2)f_1(3)\Big) \right] . \NL
 &  &
\eqry
In order to solve Eqs. (2.30) and (2.33), it is convenient to introduce
the Fourier transform, as follows:
\beq
\tilde{g}(k,p_1,p_2)=\int dq\;g(q,p_1,p_2)e^{-ik\cdot q},
\eeq
\beq
\tilde{V}(k)=\int dq\;V(q)e^{-ik\cdot q}.
\eeq
Substituting Eqs. (2.34),(2.35) into (2.30), one finds
\beq
\frac{\partial f_1}{\partial \tau }+\frac{p_1}{M}\frac{\partial f_1}{
\partial q_1}+F\frac{\partial f_1}{\partial p_1}=-\frac{i}{(2\pi )^4}\int
dk\;\left( e^{\frac{k}{2}\frac{\partial }{\partial p_1}}-e^{-\frac{k}{2}
\frac{\partial }{\partial p_1}}\right) \tilde{V}_{1,2}(k)h(k,p_1),
\eeq
where
\beq
h(k,p_1)=\int dp_2\;g(k,p_1,p_2),
\eeq
\beq
F\frac{\partial f_1}{\partial p_1}=i\left( e^{i\eta _1/2}-e^{-i\eta _1/2}
\right) f_1(1)\pm i\int dx_2\;\left( e^{i\theta ^{'}_{1,2}}-e^{-i\theta
^{'}_{1,2}}\right) f_1(1)f_1(2).
\eeq
Making a Fourier transform of Eq. (2.33), one obtains, after some
manipulations,
\bqry
{\rm Im}\;h(k,p_1) & = & \int dp_2\;\frac{\pi \tilde{V}_{1,2}(k)}{k|1\mp
\tilde{V}_{2,3}\psi |^2}\left[f_1^+(1)f_1^-(2)-f_1^+(2)f^-_1(1)\right]\NL
 &  & \times \;\delta \left( k\cdot \frac{p_1-p_2}{M}\right) .
\eqry
Here
\beq
f^{\pm}=f\left( p\pm \frac{k}{2}\right) \left[ 1\pm f\left( p\mp
\frac{k}{2}\right) \right]
\eeq
(the second sign $\pm $ in (2.40) distinguishes between bosons and
fermions),
\beq
f\left( p\pm \frac{k}{2}\right) =e^{\pm \frac{k}{2}\frac{\partial }{
\partial p}}f(p),
\eeq
and
\beq
\psi =\int _{-\infty }^\infty \frac{dp_3}{k\cdot (\frac{p_1-p_2}{M})-i
\varepsilon }\left[ f_1^+(3)-f_1^-(3)\right] .
\eeq
Substituting (2.39) into (2.36), one finally obtains
\bqry
\frac{\partial f_1(1)}{\partial \tau } & + & \frac{p_1}{M}\frac{
\partial f_1(1)}{\partial q_1}\;\;+\;\;F\frac{\partial f_1(1)}{
\partial p_1}\;\;=\;\;\frac{\pi }{(2\pi )^4}\int dk\;\left( e^{\frac{k}{
2}\frac{\partial }{\partial p_1}}-e^{-\frac{k}{2}
\frac{\partial }{\partial p_1}}\right)  \NL
 &  & \times \int dp_2\;\delta \left( k\cdot \frac{p_1-p_2}{M}\right)
\frac{\tilde{V}_{1,2}^2(k)}{|1\mp \tilde{V}_{2,3}\psi |^2}\left[ f_1^+(
1)f_1^-(2)-f_1^+(2)f_2^-(1)\right] . \NL
 &  &
\eqry
Equation (2.43) is the kinetic equation of a gas of indistinguishable
particles in the quasihomogeneous case (the improved BUU equation
\cite{YY}). It reduces to the usual BUU equation provided that the
many-body effects are neglected and that the first-order approximation
for the term $F$ is taken:
\bqry
\frac{\partial f_1(1)}{\partial \tau } & + & \frac{p_1}{M}\frac{
\partial f_1(1)}{\partial q_1}\;\;-\;\;\frac{\partial U_1}{\partial
q_1}\frac{\partial f_1(1)}{\partial p_1}\;\;=\;\;\frac{\pi }{(2\pi )^{
12}}\int dp_2dp_1^{'}dp_2^{'}\;\delta ^{4}(p_1+p_2-p_1^{'}-p_2^{'}) \NL
 &  & \times \;\Big| \langle p_1p_2|V_{1,2}|p_1^{'}p_2^{'}\rangle \Big|
^2\left\{ f_1(1^{'})f_1(2^{'})\left[ 1\pm f_1(1)\right] \left[ 1\pm
f_1(2)\right] \right.  \NL
 &  & \left. -\;\;f_1(1)f_1(2)\left[ 1\pm f_1(1^{'})\right] \left[
1\pm f_1(2^{'})\right] \right\} .
\eqry
In contrast to the usual Boltzmann and BUU equations which are applicable
in the restriction on the system to be dilute, Eq. (2.44) includes the
influence of many-body effects. Therefore, Eq. (2.44) provides an
essential improvement for the systems that have a higher particle density
or a larger force range of particle interaction; e.g., for strongly
interacting matter, heavy-ion collisions, or a cold relativistic plasma.

Rewriting this equation in the form
\bqry
\frac{\partial f_1(1)}{\partial \tau } & + & \frac{p_1}{M}\frac{
\partial f_1(1)}{\partial q_1}\;\;-\;\;\frac{\partial U_1}{\partial
q_1}\frac{\partial f_1(1)}{\partial p_1}\;\;=\;\;\frac{\pi }{(2\pi )^{
12}}\int dp_2dp_1^{'}dp_2^{'}\;\delta ^{4}(p_1+p_2-p_1^{'}-p_2^{'}) \NL
 &  & \times \;\Big| \langle p_1p_2|V_{1,2}|p_1^{'}p_2^{'}\rangle \Big|
^2\left\{ f_1(1^{'})f_1(2^{'})\left[ 1+\sigma f_1(1)\right] \left[ 1+
\sigma f_1(2)\right] \right.  \NL
 &  & \left. -\;\;f_1(1)f_1(2)\left[ 1+\sigma f_1(1^{'})\right] \left[
1+\sigma f_1(2^{'})\right] \right\} ,\;\;\;\sigma =\pm 1,
\eqry
one sees that it reduces to the usual Boltzmann equation for
non-identicai particles for $\sigma =0.$ Thus, the three cases,
$$\sigma =\left\{ \begin{array}{rl}
1, & {\rm Bose-Einstein}, \\
-1, & {\rm Fermi-Dirac}, \\
0, & {\rm Maxwell-Boltzmann},
\end{array} \right. $$ can be treated by means of a unique equation,
(2.45), which can be, therefore, called the generalized Boltzmann
equation.

\section{Boltzmann $H$-theorem}
We now wish to establish the relativistic analogue to the Boltzmann
$H$-theorem and to prove that the entropy of an ensemble of events,
evolving without external disturbances, is nondecreasing as a function of
$\tau .$

The density of states in phase space associated with the distribution $n$
has been found in \cite{HSP},
$$\triangle \Gamma (\bar{n})=\left\{ \begin{array}{ll}
(\bar{n}+g-1)!/\bar{n}!(g-1)!, & {\rm Bose-Einstein} \\
g!/\bar{n}!(g-\bar{n})!, & {\rm Fermi-Dirac} \\
g^{\bar{n}}/\bar{n}! & {\rm Maxwell-Boltzmann}
\end{array} \right. $$ where $g$ is a number of states in each elementary
cell of energy-momentum space (degeneracy) and $\bar{n}$ is the average
occupation number. Assuming no degeneracy $(g\rightarrow 1)$ and using
Stirling's approximation $$\ln N!\approx N\ln N,\;\;\;N>>1,$$ we obtain
for the density of entropy in phase space, $s,$ $$\frac{s}{k_B}\equiv
\ln \triangle \Gamma (\bar{n})=\left\{ \begin{array}{ll}
-\bar{n}\ln \bar{n}+(1+\bar{n})\ln \;(1+\bar{n}), & {\rm Bose-Einstein}
 \\ -\bar{n}\ln \bar{n}-(1-\bar{n})\ln \;(1-\bar{n}), & {\rm Fermi-Dirac}
 \\ -\bar{n}\ln \bar{n}, & {\rm Maxwell-Boltzmann}
\end{array} \right. $$
\beq
=(\sigma +\bar{n})\ln \;(1+\sigma \bar{n})-\bar{n}\ln \bar{n},\;\;\;
\sigma =\pm 1,0.
\eeq
Therefore, in the case we are considering, the entropy of the ensemble
is defined by the functional
\beq
\frac{S(\tau )}{k_B}=\int dqdp\;\Big[ (\sigma +f_1(q,p;\tau ))\ln \;(1+
\sigma f_1(q,p;\tau ))-f_1(q,p;\tau )\ln f_1(q,p;\tau )\Big] .
\eeq
Then, taking the derivative of $S(\tau )/k_B,$ using Eq. (2.45) and
integration by parts of the space-time derivatives, we obtain, after some
manipulations,
\bqry
\frac{1}{k_B}\frac{dS}{d\tau } & = & \frac{\pi }{4(2\pi )^{12}}\int dqdp_
1dp_2dp_1^{'}dp_2^{'}\delta ^4(p_1+p_2-p_1^{'}-p_2^{'})\Big| \langle
p_1p_2|V_{1,2}|p_1^{'}p_2^{'}\rangle \Big| ^2  \NL
 &  & \times f_1(q,p_1;\tau )f_1(q,p_2;\tau )f_1(q,p_1^{'};\tau )f_1(q,p_
2^{'};\tau)\left[ \left( \frac{1}{f_1(q,p_1;\tau )}+\sigma \right)
\right.  \NL
 &  & \left. \times \left( \frac{1}{f_1(q,p_2;\tau )}+\sigma \right)
\;-\;\left( \frac{1}{f_1(q,p_1^{'};\tau )}+\sigma \right)
\left( \frac{1}{f_1(q,p_2^{'};\tau )}+\sigma \right) \right]  \NL
 &  & \times \left[
\ln \left\{ \left( \frac{1}{f_1(q,p_1;\tau )}+\sigma \right)
 \left( \frac{1}{f_1(q,p_2;\tau )}+\sigma \right) \right\} \right.  \NL
 &  & -\;\left. \ln \left\{ \left( \frac{1}{f_1(q,p_1^{'};\tau )}+\sigma
\right) \left( \frac{1}{f_1(q,p_2^{'};\tau )}+\sigma \right) \right\}
\right] .
\eqry
In the derivation of (3.3) the principle of microscopic irreversibility
(e.g., detailed balance) $$\Big| \langle p_1p_2|V_{1,2}|p_1^{'}
p_2^{'}\rangle \Big| ^2dp_1dp_2=\Big| \langle p_1^{'}p_2^{'}|V_{1,2}|p_1
p_2\rangle \Big| ^2dp_1^{'}dp_2^{'}$$ and the hermiticity condition
$$\Big| \langle p_1p_2|V_{1,2}|p_1^{'}p_2^{'}\rangle \Big| ^2=\Big|
\langle p_1^{'}p_2^{'}|V_{1,2}|p_1p_2\rangle \Big| ^2$$ were used.
Since $\Big| \langle p_1p_2|V_{1,2}|p_1^{'}p_2^{'}\rangle \Big| ^2
\delta ^4(p_1+p_2-p_1^{'}-p_2^{'})\geq 0,$ and the remaining factor in
the integrand is non-negative, we obtain
\beq
\frac{dS(\tau )}{d\tau }\geq 0,
\eeq
the relativistic Boltzmann $H$-theorem.

This result implies that the entropy $S(\tau )$ is monotonically
increasing as a function of $\tau ,$ and hence the evolution of the
system, as described by the generalized relativistic Boltzmann equation,
is irreversible in $\tau ,$ but {\it not necessarily in t.} In a smooth
average sense, one can argue that the entropy must increase in $t$ as
well. The support of the distribution function in $t$ is finite at each
$\tau ;$ as $\tau $ increases, this supprort moves up the $t$-axis, since
the system as a whole moves with the free motion of the center of mass.
The entropy, according to the $H$-theorem in $\tau ,$ must therefore also
be nondecreasing, in this coarse-grained sense, in $t.$ In the
nonrelativistic limit \cite{HR} $t\rightarrow \tau ,$ $S(t)$ takes on
the usual nonrelativistic form, and the nonrelativistic $H$-theorem for
evolution in $t$ is recovered.

In the special case in which the ensemble consists of positive energy (or
negative energy) states alone, a precise $H$-theorem can be proved for
the Lyapunov function $$\frac{\tilde{S}(t)}{k_B}=\int \!d^3q\;d\tau \!
\int _{p^0>0}\!\!\!d^4p\;\frac{p^0}{M}\Big[ (\sigma +f_1(q,p;\tau ))\ln
\;(1+\sigma f_1(q,p;\tau ))-f_1(q,p;\tau )\ln f_1(q,p;\tau )\Big] ,$$ by
the application of the arguments contained in ref. \cite{HSS}.

\subsection{Relativistic four-momentum distributions}
As we have seen in the preceding section, the entropy (3.2) of a system
of events increases, according to the generalized relativistic Boltzmann
equation, monotonically in $\tau .$ It means that the momentum
distribution function monotonically approaches its equilibrium value $f_1
^{(0)}(q,p).$ The equilibrium limit is achieved when
\beq
\frac{dS(\tau )}{d\tau }=0.
\eeq
Since the integrand in (3.3) is definite, (3.5) requires that, for the
equilibrium distribution $f_1^{(0)}(q,p),$
\beq
\ln \left( \!\frac{1}{f_1^{(0)}(q,p_1)}+\sigma \!\right)
+\;\ln \left( \!\frac{1}{f_1^{(0)}(q,p_2)}+\sigma \!\right) =
\ln \left( \!\frac{1}{f_1^{(0)}(q,p_1^{'})}+\sigma \!\right)
+\;\ln \left( \!\frac{1}{f_1^{(0)}(q,p_2^{'})}+\sigma \!\right) ;
\eeq
this condition implies the vanishing of the collision term in the
generalized relativistic Boltzmann equation (2.45).

Since $p_1,p_2$ and $p_1^\prime ,\;p_2^\prime $ are the initial and
final four-momenta for any scattering process, the general solution of
(3.6) is of the form
\beq
\ln \left( \frac{1}{f_1^{(0)}(q,p)}+\sigma \right) =\chi _1(q,p)+
\chi _2(q,p)+\ldots ,
\eeq
where the $\chi _i$ exhaust all quantities for which
\beq
\chi _i(q,p_1)+\chi _i(q,p_2)
\eeq
are conserved in collisions. The quantities conserved in the sense of
(3.8) are the individual event four-momentum $p^\mu $ and mass squared $m
^2\equiv -p^2$ (the latter is asymptotically conserved in the scattering
process \cite{HL}), and a constant (the one-event ``angular
momentum'' $M^{\mu \nu }=q^\mu p^\nu -q^\nu p^\mu $ also satisfies this
requirement, but does not change the structure of the result). Hence, the
most general form of $f_1^{(0)}$ is given by \cite{HSS,di}
\beq
\ln \left( \frac{1}{f_1^{(0)}(q,p)}+\sigma \right) =-A(p-p_c)^2-B,
\;\;\;\;A=A(q),\;B=B(q),
\eeq
where $p^\mu p_{c\mu }$ is an arbitrary linear combination of the
components $p^\mu ,$ so that
\beq
f_1^{(0)}(q,p)=\frac{1}{e^{-A(p-p_c)^2-B}-\sigma }=\left\{
\begin{array}{ll}
\frac{1}{\exp \{-A(p-p_c)^2-B\}-1}, & {\rm Bose-Einstein,} \\  &  \\
\frac{1}{\exp \{-A(p-p_c)^2-B\}+1}, & {\rm Fermi-Dirac,} \\  &  \\
e^{A(p-p_c)^2+B}, & {\rm Maxwell-Boltzmann.}
\end{array} \right.
\eeq
The physical properties of the distributions (3.10) are studied in
\cite{di} for the case of non-identical particles, and in \cite{ind} for
the case of identical particles. We shall normalize these distributions
as (the physical meaning of such a normalization is manifested below):
\beq
\int dqdp\;f_1^{(0)}(q,p)=V^{(4)},
\eeq
where $V^{(4)}$ is the total four-volume occupied by the ensemble in
space-time. Let us introduce the system of the space-time densities, as
follows:
\bqry
\int dp\;f_1^{(0)}(q,p) & \equiv  & n_1^{(0)}(q), \\
\int dp_1dp_2\;f_2^{(0)}(q_1,q_2,p_1,p_2) & \equiv  &
n_2^{(0)}(q_1,q_2), \\
\int dp_1dp_2dp_3\;f_3^{(0)}(q_1,q_2,q_3,p_1,p_2,p_3) & \equiv  &
n_3^{(0)}(q_1,q_2,q_3), \;\;\;\;{\rm etc.}
\eqry
Then the one-body density, $n_1^{(0)}(q),$ is normalized, in view of
(3.11), as
\beq
\int dq\;n_1^{(0)}(q)=V^{(4)}.
\eeq
In the case of no $q$-dependence of $A$ and $B,$ Eq. (3.13) yields $n_1
^{(0)}=1.$

\section{Mean-field potential. RMS}
In the equilibrium case, Eq. (2.31) for the mean-field potential entering
the generalized Boltzmann equation (2.45), reduces to
\beq
U_1(q_1)=\frac{1}{v}\int dq_2dp_2\;f_1^{(0)}(q_2,p_2)V(q_1-q_2).
\eeq
Averaging (4.1) over the ensemble gives, through (3.11)--(3.13),
\bqry
U & \equiv  & \frac{1}{2V^{(4)}}\int dq_1dp_1\;f_1^{(0)}(q_1,p_1)U_1(
q_1)  \NL
 & = & \frac{1}{2V^{(4)}v}\int dq_1dp_1dq_2dp_2\;f_1^{(0)}(q_1,p_
1)f_1^{(0)}(q_2,p_2)V(q_1-q_2)  \NL
& \cong  & \frac{1}{2V^{(4)}v}\int dq_1dq_2\;n_2^{(0)}(q_1,q_2)V(q_1-q_
2),
\eqry
where we have used the relation $f_2\approx f_1(1)f_1(2).$

The total energy density of the ensemble is defined by
\beq
\rho =\rho _0+\rho _{int},
\eeq
where $\rho _0$ is the energy density of a free gas (no-interaction case)
calculated in refs. \cite{di,ind}, and $\rho _{int}$ is the contribution
of the interaction potential which is equal to $$\rho _{int}=N_0U,$$
$N_0$ being the particle number density per unit comoving ``proper''
three-volume $V^{(3)},$ $N_0=N/V^{(3)}.$ We now assume that for the
interacting gas $V^{(4)}/N\sim v;$ it then follows form (4.2) that
\beq
\rho =\rho _0+\frac{N^2}{2(V^{(4)})^2V^{(3)}}\int dq_1dq_2\;n_2^{(0)}
(q_1,q_2)V(q_1-q_2).
\eeq
For a quasihomogeneous system, $n_2^{(0)}(q_1,q_2)=n_2^{(0)}(q_1-q_2),$
so that (4.4) takes on the form
\beq
\rho =\rho _0+\frac{N^2}{2V^{(4)}V^{(3)}}\int dq\;n_2^{(0)}(q)V(q).
\eeq
By introducing hyperbolic variables for spacelike $q,$
$$q^0=q\sinh \beta ,\;\;\;q^1=q\cosh \beta \sin \theta \cos \phi ,$$
\beq
q^2=q\cosh \beta \sin \theta \sin \phi ,\;\;\;q^3=q\cosh \beta \cos
\theta ,
\eeq
$$0\leq q<\infty ,\;\;-\infty <\beta <\infty ,\;\;0\leq \theta \leq \pi ,
\;\;0\leq \phi <2\pi ,$$ one can rewrite the integral in Eq. (4.5) as
$$4\pi \int q^3dq\cosh ^2\beta d\beta \;n_2^{(0)}(q)V(q).$$ This integral
does not have, however, a simply interpretable nonrelativistic limit, as
we discuss below after Eq. (4.12).
Let us instead turn to ref. \cite{AH1}, where the two-body
relativistic quantum-mechanical bound-state problem has been studied. It
was found that, if the support of the wave function of the relative
motion is restricted to an $O(2,1)$-invariant subregion of the full
spacelike region, one finds a lower mass eigenvalue of the ground state
than in the case when the support is in the full spacelike region. The
solutions, moreover, have a simply interpretable nonrelativistic limit.
This subregion was called by the authors the ``restricted Minkowski
space'' (RMS). It has a parametrization (in contrast to (4.6)
corresponding to the full spacelike region) $$q^0=q\sin \theta \sinh
\beta ,\;\;\;q^1=q\sin \theta \cosh \beta \cos \phi ,$$
\beq
q^2=q\sin \theta \cosh \beta \sin \phi ,\;\;\;q^3=q\cos \theta .
\eeq
Clearly, $q_1^2+q_2^2-q_0^2=q^2\sin ^2\theta \geq 0$ (and $q_1^2+q_2^2+
q_3^2-q_0^2=q^2\geq 0$ as well). This submeasure space is
$O(2,1)$-invariant, but not $O(3,1)$-invariant. The representations of
$O(3,1)$ are induced from the irreducible representations of $O(2,1)$
which are provided by the eigenfunctions of the two-body bound-state
problem \cite{AH2}. The fact that this restricted subregion admits a
lower mass of the ground state than the full spacelike region constitutes
a spontaneous symmetry breaking of the $O(3,1)$ invariance of the
dynamical equations.

The restriction of the relative coordinates to the RMS corresponds to a
restricted range of correlations available to the two events
propagating in a bound state, i.e., to the range of $q^\mu _1-q^\mu _2$
available at each $\tau .$ In computing the full spectrum of the two-body
problem the authors assumed that the wave functions of the excited states
also lie in the $O(2,1)$-invariant subregion, i.e., these correlations
are maintained for excited states as well. Indeed, it was found that the
partial wave expansion for scattering theory is recovered in this
submeasure space as well \cite{AH3}. Here we shall assume that this
result has more generality and can be applied in statistical mechanics:
{\it for any two events, their mutual correlations lie in the relative}
$O(2,1)$-{\it invariant subregion of the full spacelike region.}
It then follows that the two-body density $n_2^{(0)}(q_1-q_2)$ will have
support lying in the RMS associated with the relative motion $q^\mu _1-q^
\mu _2.$ Therefore, the integral in Eq. (4.5) will be nonzero only in
the RMS associated with $q,$ according to the nonvanishing support of the
two-body density $n_2^{(0)}(q).$ In this way we obtain for the integral
in Eq. (4.5)
\beq
\int q^3dq\sin ^2\theta d\theta \cosh \beta d\beta d\phi \;n_2^{(0)}(q)
V(q).
\eeq
We shall also assume that, at any instant of $\tau ,$ the extent of the
ensemble in the $q^0$-direction is bounded \cite{HSP}, so that $V^{(4)}=
V^{(3)}\cdot \triangle t,$ where $\triangle t$ is the range of the time
variable for the system as a whole. Therefore, in Eq. (4.8) $-\frac{
\triangle t}{2}\leq q^0\leq \frac{\triangle t}{2},$ and integration on
$\beta $ gives $$\int _{-{\rm Arc}\!\sinh (\triangle t/2q\sin \theta )}^
{{\rm Arc}\!\sinh (\triangle t/2q\sin \theta )}\cosh \beta d\beta =
\frac{\triangle t}{q\sin \theta };$$ Eq. (4.8) then reduces to
\beq
\triangle t\int q^2dq\sin \theta d\theta d\phi \;n_2^{(0)}(q)V(q)=
4\pi \triangle t\int dq\;q^2n_2^{(0)}(q)V(q).
\eeq
Using now the relations $V^{(4)}=V^{(3)}\cdot \triangle t$ and $N/V^{(3)}
=N_0,$ the particle number density, one finally obtains from (4.5),(4.9)
\beq
\rho =\rho _0+\frac{N_0^2}{2}\int d^3q\;n_2^{(0)}(q)V(q),
\eeq
where $d^3q$ stands for $4\pi q^2dq.$
In the same way it is possible to obtain the expression for the pressure
of the interacting gas \cite{therm}:
\beq
p=p_0-\frac{N_0^2}{6}\int d^3q\;q\frac{dV(q)}{dq}n_2^{(0)}(q).
\eeq
We see that the expressions for $\rho $ and $p$ are precisely of the same
form as those of nonrelativistic statistical mechanics \cite{Fey}, but
with $q\equiv \sqrt{q^\mu q_\mu }$ replacing $r\equiv \sqrt{{\bf q}^2},$
and $V(q)$ replacing $V(r).$ The situation is quite similar to the one
occuring in the two-body bound-state problem \cite{AH1}, where, upon
separation of variables in the RMS, one is left with a radial equation
for $q\equiv \sqrt{q^\mu q_\mu }$ which is of the same form as a
nonrelativistic radial Schr\"{o}donger equation for $r\equiv \sqrt{{\bf q
}^2}.$ Separation of variables in the RMS therefore has a clear
correspondence to the nonrelativistic problem, as first remarked in
\cite{AH1}. In the nonrelativistic limit, the relative variables $q^0$
and $p^0$ vanish (all the particles are synchronized in this limit
\cite{HSS}), and the formulas (4.10),(4.11) acquire their standard
nonrelativistic expressions \cite{Fey}.

We remark that the integral in Eq. (4.5) in the full spacelike region
can be made convergent in the same way, by imposing the bounds on the
time variable, as follows: $-\triangle t/2\leq q\sinh \beta \leq
\triangle t/2.$ In this case integration on $\beta $ results in the
expression $$4\pi \left( \frac{\triangle t}{2}\right) ^2\int dq\;
qV(q)n_2^{(0)}(q),$$ and we obtain, in place of (4.10),
\beq
\rho =\rho _0+\frac{N_0^2}{2}\;\frac{\triangle t}{4}\;
4\pi \int dq\;q\;n_2^{(0)}(q)V(q),
\eeq
and similar relation for $p.$ Hence, apart from $T_{\triangle V},$ the
average passage interval in $\tau $ for the events which pass through a
small representative four-volume of the system \cite{HSS}, contained in
the expressions for $\rho _0,p_0$ and $N_0$ \cite{di,ind}, there will
be another $T_{\triangle V}$ entering the expressions for $\rho _{int}$
and $p_{int},$ upon replacement for $\triangle t$ in the corresponding
formulas, through the relation (which represents the averaging of the
equation of motion for $q^0,$ $dq^0/d\tau =p^0/M,$ over the ensemble,
$\langle E\rangle $ being the average energy) $$\triangle t=T_{\triangle
V}\frac{\langle E\rangle }{M}.$$ In the nonrelativistic (or in the
sharp-mass) limit, $T_{\triangle V}\rightarrow \infty ,$ which provides a
stationarity of the system in space-time, but not a non-trivial evolution
in $\tau $ \cite{jpa}. While $p_0,\rho _0$ and $N_0$ are preserved in
this singular limit, due to the relation \cite{jpa} $T_{\triangle V}
\triangle m=2\pi ,$ where $\triangle m$ is the width of the mass
deviation from its on-shell value, $p_{int}$ and $\rho _{int}$ turn out
to converge with $T_{\triangle V}.$ Therefore, Eq. (4.12) and similar
formula for $p$ do not have a well-defined nonrelativistic limit, in
contrast to (4.10),(4.11), which admit its clear form. This fact should
be a source of a spontaneous symmetry breaking of the $O(3,1)$ invariance
in the correlation function of a many-body problem.

We remark that no problem with the convergence of the integral in Eq.
(4.5) arises in 1+2 dimensions (for the extent in the $q^0$-direction
bounded). Indeed, the 3D analog of (4.5) reads
\beq
\rho =\rho _0+\frac{N^2}{2V^{(3)}V^{(2)}}\int d^3q\;n_2^{(0)}(q)V(q).
\eeq
By introducing hyperbolic variables for spacelike $q,$
\beq
q^0=q\sinh \beta ,\;\;\;q^1=q\cosh \beta \sin \theta ,\;\;\;
q^2=q\cosh \beta \cos \theta ,
\eeq
$$0\leq q<\infty ,\;\;-\infty <\beta <\infty ,\;\;0\leq \theta \leq
2\pi ,$$ we rewrite the latter integral as $$2\pi \int q^2dq\cosh \beta d
\beta \;n_2^{(0)}(q)V(q).$$ Integration on $\beta $ gives $$\int _{-{\rm
Arc}\!\sinh (\triangle t/2q)}^{{\rm Arc}\!\sinh (\triangle t/2q)}\cosh
\beta d\beta =\frac{\triangle t}{q};$$ therefore, one obtains, via
$\triangle t=V^{(3)}/V^{(2)},$ $N/V^{(2)}=N_0,$
\beq
\rho =\rho _0+\frac{N_0^2}{2}\int d^2q\;n_2^{(0)}(q)V(q),
\eeq
the 3D analog of (4.10) ($d^2q$ stands for $2\pi qdq),$ and a similar
relation for $p.$

\section{Concluding remarks}
We have generalized the nonrelativistic approach of Yang and Yao, based
on the Wigner distribution functions and the Bogoliubov hypotheses, to
the relativistic case. We have derived the generalized Boltzmann equation
which, in the case of indistinguishable particles, improves the standard
BUU equation in three main aspects:

1) The effect of Pauli blocking, $f\rightarrow f(1-f^{'}),$ is included
in the collision term. This is important for the collision processes at
intermediate and low temperature, e.g., in heavy-ion collisions.

2) The modified mean-field interaction is introduced into the collision
term. This has a great influence on far-nonequilibrium states.

3) The equation takes into account binary collisions corrected for
many-body effects, wherein the many-body shielding effect can be obtained
spontaneously.

We have introduced two-body interactions, by means of the direct action
potential $V(q),$ where $q$ is an invariant distance in the Minkowski
space. The two-body correlations are taken to have the support in a
relative $O(2,1)$-invariant subregion of the full spacelike region, in
order to provide a good nonrelativistic limit to the basic thermodynamic
quantities. Since the expressions for the energy density and the pressure
are identical in form to those of the nonrelativistic theory, some of the
results for the nonrelativistic interacting gas should be applicable for
an interacting off-shell gas as well. For example, the equation of state
of the ideal gas of non-identical particles is \cite{di} $p=N_0T;$
therefore, it follows from (4.10),(4.11) that the equation of state of a
relativistic interacting gas should have the same form (by methods
analogous to those of the standard cluster expansion \cite{Hua}) as that
of a similarly interacting nonrelativistic one, i.e. \cite{Hua},
$$\frac{p}{N_0T}=\sum _{l=1}^\infty a_l(T)\left( \lambda ^3N_0\right) ^{
l-1},$$ where $\lambda \equiv \sqrt{2\pi /MT}$ is the thermal
wavelength, and $a_l(T)$ is the $l$th virial coefficient ($a_1=1).$

Applications of the generalized Boltzmann equation to realistic physical
systems, e.g., heavy-ion collisions, are now under consideration.

\section*{Acknowledgments}
We wish to thank R.I. Arshansky for discussions concerning the notion of
the restricted Minkowski space (RMS) and its role in relativistic
physics.

\end{document}